\documentclass[preprint,showpacs,amsmath,amssymb]{revtex4}
\usepackage{amssymb}

\usepackage{graphicx}
\usepackage{dcolumn}
\usepackage{bm}

\begin{document}

\title{Contribution of the t-channel $N^*(1535)$ exchange for the $p\bar p \to \phi\phi$ reaction}

\author{Jun Shi$^{1,2}$}
\author{Jian-Ping Dai$^{1}$}
\author{Bing-Song Zou$^{1,3}$}

\affiliation{$^1$ Institute of High Energy Physics and Theoretical
Physics Center for Sciences Facilities, CAS,
Beijing 100049, China\\
$^2$ Shandong university at Weihai, Weihai 264209, China\\
$^3$ Center of Theoretical Nuclear Physics, National Laboratory of
Heavy Ion Collisions, Lanzhou 730000, China}

\begin{abstract}
Since the $N^*(1535)$ resonance was found to have large coupling to
the strangeness due to its possible large $s\bar s$ component, we
investigate the possible contribution of the t-channel $N^*(1535)$
exchange for the $p\bar p \to \phi\phi$ reaction. Our calculation
indicates that the new mechanism gives very significant contribution
for the energies above 2.25 GeV and may be an important source for
evading the Okubo-Zweig-Iizuka rule in the $\phi$ production from $N
\bar{N}$ annihilation.
\end{abstract}
\pacs {13.75.-n, 13.75.Cs, 14.20.Gk, 25.75.Dw}
\maketitle{}

\section{INTRODUCTION}

$\phi$ meson production from $N \bar{N}$ annihilation has been
drawing intensive interest for its violation of the Okubo-Zweig-Iizuka
(OZI) rule~\cite{ozi,zouozi,nomo}. The usual statement of the OZI rule
is that diagrams with disconnected quark lines can be negligible
compared to those with connected quark lines. $\phi$ meson is
believed to be an almost pure $s \overline{s}$ state, while nucleon
is universally considered to be composed of up and down quarks. Thus
according to the OZI rule, the reaction that $N \bar{N}$ annihilate to
produce $\phi$ is to be suppressed. However, it is known that $\phi$
production evades the OZI rule in various hadronic
reactions~\cite{exp}. The study of the reaction is seen as a
promising probe into the strangeness information of nucleon or
nucleon resonances.

To interpret the substantial OZI rule violations,
Lindenbaum~\cite{glueball0} and Etkin et al.~\cite{glueball2}
pointed out the intervention of glueball resonances,
Dover~\cite{fourquark} argued $s \overline{s} n \overline{n}$ four
quark states produce selective enhancements,
Kochelev~\cite{instanton} took instanton effects into account, Ellis
et al.~\cite{nucleon1,nucleon2} inferred considerable admixture of
$s \overline{s}$ components in the nucleon. Also, as OZI-forbidden
reactions can proceed via two-step hadronic loops in which each
individual transition is OZI-allowed~\cite{Lipkin}, Lu et
al.~\cite{twomeson} studied two-meson ($K \overline{K}$)
intermediate states and Mull et al.~\cite{twohyperon} investigated
antihyperon-hyperon ($\Lambda \overline{\Lambda}$) intermediate
states, both of them educed compatible theoretical results with the
experimental data.

On the other hand, recently, it has been suggested that the strong
couplings of $N^*(1535)$ to $\eta N$, $\eta' N$ and $K \Lambda$ may
indicate a possible significant $s \bar{s}$ component in the quark
wave function of $N^*(1535)$ and also a large coupling to $\phi
N$~\cite{xiejujun,caoxu,oset}. Xie et al.~\cite{xiejujun} deduced
the significant coupling of $N^*(1535)$ to $N \phi$ as
$g_{N^*(1535)N \phi}^2/4\pi=0.13$. With such effective $N^*(1535)N
\phi$ coupling, the measured $\pi^- p\to n\phi$, $pp\to pp\phi$ and
$pn\to\phi d$ cross sections are well
reproduced~\cite{xiejujun,caoxu}.

In this article, we extend the model used in Ref.~\cite{xiejujun} to
study the $p \bar{p}\to\phi\phi$ reaction. The scenario is that $p
\bar{p}$ interact with each other by exchanging $N^*(1535)$ in $t$
channel, which has not been attempted by others so far.

\section{FORMALISM AND INGREDIENTS} \label{formalism}

\begin{figure}[htbp]
  \begin{center}
\begin{minipage}[t]{70mm}
 {\includegraphics*[scale=0.7]{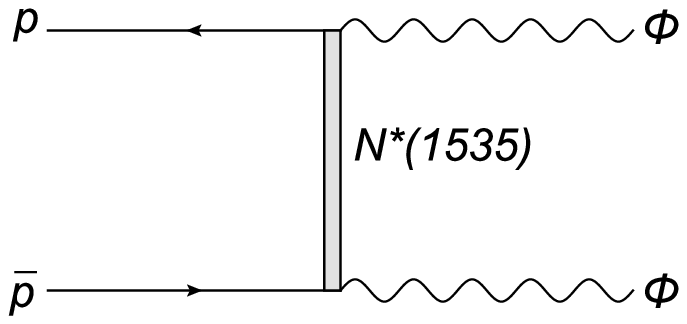}}\vskip -0cm
\end{minipage}
\begin{minipage}[t]{70mm}
{\includegraphics*[scale=0.7]{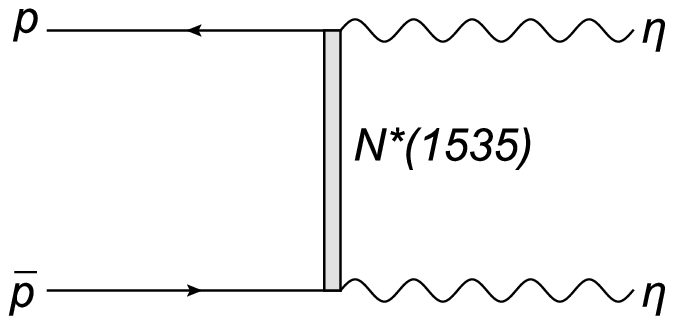}}\vskip -0cm
\end{minipage}
\caption{Feynman diagrams for $p \bar{p}\to\phi\phi$ and $\eta\eta$
with $t$-channel $N^*(1535)$ exchange}
  \end{center}
\end{figure}

The Feynman diagram for the $p \bar{p}\to\phi\phi$ reaction is
depicted in Fig.1~(left). We use the effective Lagrangian for
$N^*(1535)N \phi$ as~\cite{xiejujun}
\begin{equation}
{\cal L}_{\phi N N^*}= i g_{N^* N \phi}\overline{N} \gamma_5
(\gamma^{\mu}-\frac{q^{\mu} \not \! q}{q^2})\phi_{\mu}N^*+h.c..
\end{equation}
Here $N$ and $N^*$ are the spin wave functions for the nucleon and
$N^*(1535)$ resonance, respectively, and $\phi_{\mu}$ is the
$\phi$-meson field. For the $N^*(1535)$-$N$-$\phi$ vertices, the
following monopole form factor is used:
\begin{equation}
F_{N^*}(q^2)=\frac{\Lambda^2-m_{N^*}^2}{\Lambda^2-q^2}.
\end{equation}
Empirically the cut-off parameter $\Lambda$ for $N^*(1535)$ should
be at least a few hundred MeV larger than the $N^*$ mass, hence to
be in the range of 2 to 4 GeV.

Then the amplitude and cross section can be obtained
straightforwardly by applying the Feynman rules to Fig.1~(left).
\begin{equation}
\begin{aligned}
{\mathcal M}=&-g_{N^*N
\phi}^2F_{N^*}^2(q^2){\overline{v}}_{\overline{p}}(p',s')\gamma_5
(\gamma^{\mu}-\frac{q^{\mu} \not \!
q}{q^2}){\epsilon}_{\mu}^*(k')\frac{i}{\not \! q -M_{N^*}}\gamma_5
(\gamma^{\nu}-\frac{q^{\nu} \not \!
q}{q^2}){\epsilon}_{\nu}^*(k)u_{p}(p,s) \\
&+(\texttt{exchange term with } k\leftrightarrow k'),
\end{aligned}
\end{equation}
with $q=p-k$, and
\begin{equation}
\frac{d \sigma}{d \Omega}=\frac{\left|\vec{k}\right|}{256{\pi}^2
\left|\vec{p}\right|(p+p')^2}\sum\limits_{s,s'}\left|{\mathcal
M}\right|^2.
\end{equation}

\section{Numerical RESULTS AND DISCUSSION} \label{discussion}

\begin{figure}[htbp]
  \begin{center}
 {\includegraphics*[scale=0.8]{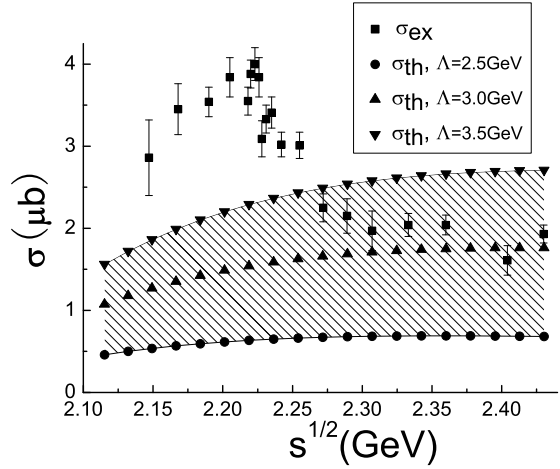}}
\caption{Theoretical results for the contribution from the t-channel
$N^*(1535)$ exchange to the $p \bar{p}\to\phi\phi$ reaction with
various $\Lambda$ parameter, compared with data~\cite{data}.}
  \end{center}
\end{figure}

Fig.2 shows the theoretical results for the contribution from the
t-channel $N^*(1535)$ exchange to the $p \bar{p}\to\phi\phi$
reaction with various $\Lambda$ parameter, compared with the
experimental data~\cite{data}. One can see that for the invariant
mass $s^{1/2}$ above 2.3 GeV the new mechanism alone reproduces the
data well with $\Lambda\sim 3$ GeV.

To check whether the new mechanism and its $\Lambda$ parameter are
reasonable, we also use the same approach to study $p \bar{p}
\to\eta\eta$ as shown in Fig.1~(right). The interaction Lagrangian
for $N^*(1535)N\eta$ coupling is~\cite{xiejujun}:
\begin{equation}
{\cal L}_{\eta N N^*(1535)}=i g_{N^* N \eta}\overline{N} \eta
N^*+h.c.,
\end{equation}
with $g_{\eta N N^*(1535)}/4\pi=0.28$~\cite{xiejujun}. The
theoretical results for the contribution from the t-channel
$N^*(1535)$ exchange with $\Lambda$=3 and 3.5 $\rm{GeV}$ are shown
in Fig.3. Compared with experimental data~\cite{etaex}, the
contribution is very small. This is consistent with the partial wave
analysis of Ref.\cite{etapwa} which suggests a large s-channel
resonant contribution from $0^{++}$, $2^{++}$ and $4^{++}$ meson
resonances with masses of $2.0\sim 2.4$ GeV. These meson resonances
have strong couplings to both $p\bar p$ and $\eta\eta$, allowed by
the OZI rule. Hence the t-channel $N^*(1535)$ exchange only plays a
minor role for the $p \bar{p} \to\eta\eta$ reaction. These meson
resonances can also decay to $\phi\phi$ through strange meson or
baryon loops~\cite{twomeson,twohyperon}, which give a few times
smaller rate. Then the t-channel $N^*(1535)$ exchange plays a more
important role for the $p \bar{p} \to\phi\phi$ reaction. However,
while the contribution from the t-channel $N^*(1535)$ exchange can
reproduce the $p \bar{p} \to\phi\phi$ cross sections above 2.3 GeV,
it cannot reproduce the peak structure around 2.2 GeV. This suggests
that the s-channel resonances may still dominates here through the
strange meson or baryon loops.

\begin{figure}[htbp]
  \begin{center}
 {\includegraphics*[scale=0.8]{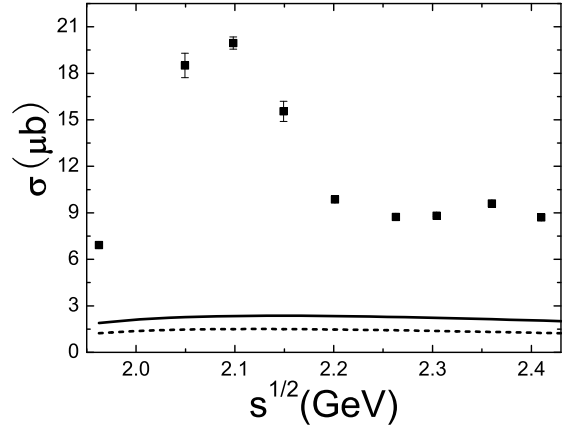}}
\caption{Cross sections for $p \bar{p}\to\eta\eta$ with cos$\theta$
integrated from 0 to 0.85~\cite{etaex}. The dashed and solid curves
show the contribution from the t-channel $N^*(1535)$ exchange with
$\Lambda$=3 and 3.5 $\rm{GeV}$, respectively. }
  \end{center}
\end{figure}

In summary, we have phenomenologically studied the contribution of
the t-channel $N^*(1535)$ resonance exchange to the $p \bar{p}
\to\phi\phi$ and $\eta\eta$ reactions. While it plays only a minor
role for the $p \bar{p} \to\eta\eta$ reaction, it gives a
significant contribution to the $p \bar{p} \to\phi\phi$ reaction,
especially for the energies above 2.3 GeV. Below 2.3 GeV, the
s-channel resonances may still dominates through the strange meson
or baryon loops. Before one tries to extract any definite
information about the strangeness of the proton from this reaction
as suggested by Ref.\cite{nucleon2}, one should take these
additional mechanisms into account.

\begin{acknowledgments}

Useful discussions with Xu Cao, Jia-Jun Wu, Lu Zhao and Pu-Ze Gao
are gratefully acknowledged. This work is supported by the National
Natural Science Foundation of China (NSFC) under grants Nos.
10875133, 10821063, and by the Ministry of Science and Technology of
China (2009CB825200).

\end{acknowledgments}


\begin{thebibliography}{1}
%
\bibitem {ozi}S. Okubo, Phys. Lett. B \textbf{5}, 165 (1963);
G. Zweig, CERN Report No. 8419/TH412, 1964; J. Iizuku, Prog. Theor.
Phys. Suppl. \textbf{38}, 21 (1996).
%
\bibitem {zouozi} B. S. Zou, Phys. Atom. Nucl. \textbf{59}, 1427
(1996).
%
\bibitem {nomo} V.P. Nomokonov and M.G. Sapozhnikov, Phys. Part.
Nucl. \textbf{34}, 94 (2003).
%
\bibitem {exp} J. Ellis, E. Gabathuler and M. Karkiner, Phys. Lett.
B \textbf{217} 173 (1989); C. Amsler et al., Crystal Barrel
Collaboration, Phys. Lett. B \textbf{346}, 363 (1995); M.G.
Sapozhnikov, Nucl. Phys. A \textbf{655} 151 (1999).
%
\bibitem {glueball0} S. J. Lindenbaum, Nuovo Cim. A \textbf{65}, 222
(1981).
%
\bibitem {glueball2} A. Etkin et al. , Phys. Lett. B \textbf{201},
568 (1988).
%
\bibitem {fourquark} C. B. Dover, P. M. Fishbane, Phys. Rev.
Lett. \textbf{62}, 2917 (1989).
%
\bibitem {instanton} N. I. Kochelev, Phys. Atom. Nucl.
\textbf{59},1643 (1996).
%
\bibitem {nucleon1} J. R. Ellis, E. Gabathuler, M. Karliner,
Phys. Lett. B \textbf{217}, 173 (1989).
%
\bibitem {nucleon2} J. R. Ellis, M. Karliner,  D. E. Kharzeev,  M. G.
Sapozhnikov, Phys. Lett. B \textbf{353}, 319 (1995); Nucl. Phys. A
\textbf{673}, 256 (2000).
%
\bibitem {Lipkin} H. J. Lipkin, Nucl. Phys. B \textbf{244}, 147 (1984);  \textbf{291}, 720
(1987).
%
\bibitem {twomeson} Y. Lu, B.S. Zou, M.P. Locher, Z. Phys. A
\textbf{345}, 207 (1993).
%
\bibitem {twohyperon} V. Mull, K. Holinde, J. Speth, Phys. Lett. B
\textbf{334}, 295 (1994).
%
\bibitem {xiejujun} J. J. Xie, B. S. Zou, H. C. Chiang,
Phys. Rev. C \textbf{77}, 015206 (2008).
%
\bibitem {caoxu} Xu Cao, J. J. Xie, B. S. Zou, H. S. Xu,
Phys. Rev. C \textbf{80}, 025203(2009).
%
\bibitem{oset} M.~Doring, E.~Oset and B.~S.~Zou,
  Phys.\ Rev.\  C {\bf 78}, 025207 (2008).
%
%
\bibitem {data} C. Evangelista et al. , JETSET Collaboration, Phys.
Rev. D \textbf{57}, 5370 (1998).
%
\bibitem {etaex} A.V. Anisovich et al., Nucl. Phys. A \textbf{662}, 344 (2000).
\bibitem{etapwa} A.V. Anisovich et al., Nucl. Phys. A \textbf{662}, 319 (2000).
%
\end{thebibliography}
\end{document}